\documentclass[12pt]{iopart}

\usepackage{subfigure}
\usepackage{graphicx}
\usepackage{epsfig}
\usepackage[small]{caption}
\begin{document}

\title[The nonlinear dielectric waveguide -- surface-plasmon Josephson Junction]{Dynamical analysis of a weakly coupled nonlinear dielectric waveguide -- surface-plasmon model as a new type of Josephson Junction}

\author{Yasa Ek\c{s}io\u{g}lu,$^1$ \"{O}zg\"{u}r E. M\"{u}stecapl{\i}o\u{g}lu,$^1$ and Kaan G\"{u}ven$^{1}$}

\address{$^1$Department of Physics, Ko\c{c} University, Istanbul\ Turkey}

\ead{yeksioglu@ku.edu.tr}

\begin{abstract}
We propose that a weakly-coupled nonlinear dielectric waveguide -- surface-plasmon system can be formulated as a new type of Josephson junction. Such a system can be realized along a metal - dielectric interface where the dielectric medium hosts a nonlinear waveguide (e.g. fiber) for soliton propagation. We demonstrate that the system is in close analogy to the bosonic Josephson-Junction (BJJ) of atomic condensates at very low temperatures, yet exhibits different dynamical features.  In particular, the inherently dynamic coupling parameter between soliton and surface-plasmon generates self-trapped oscillatory states at nonzero fractional populations with zero and $\pi$ time averaged phase difference. The salient features of the dynamics are presented in the phase space.

\end{abstract}


\section{Introduction}
\label{sec:Introduction}

Plasmonics encompasses the science and technology of plasmons which are the collective oscillations of electrons mainly in metals. In particular, the coupling of plasmons with photons give rise to a hybrid quasiparticle known as the surface plasmon-polariton (also called surface-plasmon)~\cite{proposal8}, which can propagate along metal surfaces. As a result, light can be coupled to, and propagate through structures that are much smaller than it's wavelength. Thus, plasmonics enables the subwavelength photonics~\cite{proposal50,proposal64} and bridges it to electronics at nanoscale and offers promising applications in nano-optics and electronics such as lasing, sensing \cite{maier,surface-plasmon-nanophotonics,ritchie,ybliokh}.\\

In this context, the coupling between the surface plasmon polaritons and the dielectric waveguide modes is of interest for  integrated optoelectronic applications. Much effort has been devoted to understand and enhance this coupling in guided mode geometries~\cite{Hochberg2004,FangLiu2007,Zia2005} and in layered systems~\cite{Ditlbacher2008}. In particular, a recent work describes the resonant interaction model between the surface-plasmons on a metal surface and a soliton in a nonlinear dielectric medium.~\cite{proposal1} In that model, the interaction depends on, and hence can be controlled by the soliton amplitude which leads to coupled surface-plasmon--soliton modes, and provides a novel way of manipulating the surface plasmon propagation. Motivated by this study, we propose in this paper that a weakly coupled nonlinear dielectric waveguide surface-plasmon system can be formulated as a new type of `dielectric waveguide -- surface-plasmon Josephson Junction (DWSP-JJ)' and investigate its dynamics within the model introduced in Ref.~\cite{proposal1}. This formulation enables a remarkable connection between this optical system and the well-known Josephson-Junction dynamics of Bose-Einstein atomic condensates~\cite{raghavan} and opens a new perspective to quantum plasmonics, where novel quantum optical phenomena can be realized and utilized via coupling to surface plasmons.~\cite{chang2006}

The paper is organized as follows. In Section \ref{sec:Theory}, the theoretical model is introduced. Section \ref{sec:Results}, investigates the dynamical features of the DWSP-JJ system in comparison to SJJ and BJJ systems. Section \ref{sec:Conclusion} concludes the paper.

\section{Model and Theoretical Formulation}
\label{sec:Theory}

Within the scope of this paper, we adopt the model system given in Ref.~\cite{proposal1}. This intuitive model consists of the co-propagating optical soliton and surface-plasmon (SP) electric fields along a metal-dielectric interface~\cite{plasmon-soliton,self}. The nonlinearity in the dielectric is assumed to be confined at a distance, $d$, from the metal interface such that the surface-plasmon propagation retains linearity, and the weak coupling between the soliton and SP-fields can be treated perturbatively. Assuming translational invariance in the $z$ direction, and choosing $y$ as the propagation direction, the total electric field of the coupled system is written in the form,
\begin{equation}
\ E(x,y) = c_p(y)e^{-\kappa_px} + \frac{c_s(y)}{\cosh\left[\kappa_s(x-d)\right]},
\label{eq:TotalField}
\end{equation}
where $c_{p,s}(y)$ are the surface-plasmon and soliton amplitudes, multiplied by the respective transverse profiles with $\kappa_p=\sqrt{k_p^2-k^2}$ and $\kappa_s=k\sqrt{\gamma/2}|c_s|$. The propagation wavevector is $k$ whereas $k_p$ is the surface-plasmon wavevector. The nonlinearity parameter of the dielectric medium is $\gamma$. This product ansatz for the lateral and longitudinal field profiles reduces the system to essentially a one-dimensional problem, where the amplitudes $c_{p,s}(y)$ obey the following coupled oscillator equations, that are written in the dimensionless coordinate $\zeta = ky$

\begin{equation}
\begin{array}{ll}
\ \ddot{C_p} + \beta_p^2c_p = q(|c_s|)c_s,\\
\ \ddot{C_s} + \beta_s^2c_s = q(|c_s|)c_p.
\end{array}
\label{eq:CoupledOscillator}
\end{equation}

Here, $\beta_p = k_p/k$ and $\beta_s=1+\gamma|c_s|^2/4$ are the propagation constants. The coupling parameter is given by $q(|c_s|)\simeq \exp(-k\sqrt{\gamma/2}|c_s|d)$ \cite {proposal1}. The rationale of this functional form is that the lateral tail of the soliton field, which depends on the soliton amplitude $|c_s|$, acts as a source to excite the surface plasmons. Substituting $c_{p,s} = C_{p,s}e^{i\zeta}$ and employing the slowly varying amplitude approximation, the following set of equations are obtained for the amplitudes $C_{p,s}$

\begin{equation}
\begin{array}{ll}
\ -i\dot{C_p}=\nu_pC_p-\frac{q(|C_s|)}{2}C_p,\\
\ -i\dot{C_s}=-\frac{q(|C_s|)}{2}C_p+\nu_s(|C_s|)C_s.
\end{array}
\label{eq:Cp_and_Cs}
\end{equation}

In Eq.~\ref{eq:Cp_and_Cs}, $\nu_p\equiv\beta_p-1\ll1$, $\nu_s\equiv\beta_s-1\ll1$ are the small deviations of the dimensionless propagation constants of surface-plasmon and soliton, respectively.
The eigenmode analysis indicates that the resonant coupling $q_{res} = \exp[-kd\sqrt{2\nu_p}]$ occurs around $\nu_p=0.2$, $\eta=0.2$ ~\cite{proposal1}.

\subsection{Josephson Junction Formulation}
Remarkably, the Eq.~\ref{eq:Cp_and_Cs} can be cast in analogy to the Josephson junction dynamics by writing $C_{s,p} = \mathcal{C}_{s,p}e^{i\phi_{s,p}}$, and introducing the fractional population imbalance $Z = (|C_s|^2 - |C_p|^2)/N$ and the relative phase difference between the soliton and the surface plasmon $\phi = \phi_s - \phi_p$ as follows:
\begin{equation}
\ \dot{Z}=-q(Z)\sqrt{1-Z^2}\sin\phi,
\label{eq:DWSP_JJ_Z}
\end{equation}
\begin{equation}
\ \dot{\phi}= \Lambda Z+\triangle E+\frac{q(Z)Z}{\sqrt{1-Z^2}}\cos\phi.
\label{eq:DWSP_JJ_phi}
\end{equation}
We set $N = (|C_s|^2+|C_p|^2)\equiv 1$, a normalized constant for the isolated system with no population dissipation, and define the parameters $\Lambda\equiv\eta/2$, $\triangle E\equiv\eta/2-\nu_p$, with $\eta=\frac{\gamma N}{4}$. $\Lambda$ is the nonlinearity (i.e. the soliton strength) whereas $\triangle E$ parametrizes the asymmetry between the soliton and surface plasmon states occupied by the photons.The coupling parameter takes the $Z$-dependent form
\begin{equation}
\ q(Z)=e^{-kd\sqrt{2\Lambda(1+Z)}}.
\label{eq:q}
\end{equation}

An immediate comparison reveals the similarities and differences to the dynamical model of the bosonic Josephson junction (BJJ) in a double well trap model of two Bose-Einstein condensates~\cite{raghavan, hamiltonian}:

\begin{equation}
\ \dot{Z}=-\sqrt{1-Z^2}\sin\phi,
\label{eq:BJJ_Z}
\end{equation}
\begin{equation}
\ \dot{\phi}= \Lambda Z+\triangle E+\frac{Z}{\sqrt{1-Z^2}}\cos\phi
\label{eq:BJJ_phi}
\end{equation}
In the BJJ model, $\Lambda$ describes the interatomic interactions and $\triangle E$ is the difference between the zero-point energies of the trapping wells. Both $\Lambda$ and $\triangle E$ are dimensionless parameters scaled by the coupling matrix element.~\cite{hamiltonian} We observe that the DWSP-JJ equations would reduce to that of the BJJ model for a constant coupling parameter $q(Z)\equiv 1$ provided that the other parameters are scaled accordingly.

It is instructive to present also the case of superconducting Josephson junctions (SJJ) where the dynamical variable is the voltage across the resistive shunt between the two superconductors~\cite{bdjosephson}:
\begin{equation}
\ \dot{\phi} =\frac{2eV(t)}{\hbar}
\label{eq:SJJ}
\end{equation}
The SJJ tunneling AC current is given by,
\begin{equation}
\ \dot{I} = I_0 (\cos\phi)\dot{\phi}.
\label{eq:SJJ_current}
\end{equation}
The Cooper-pair population imbalance across SJJ is zero when the SJJ is closed over an external circuit~\cite{josephson}. In the absence of external circuit, an isolated SJJ can exhibit coherent Cooper-pair oscillations only at very small amplitudes~\cite{josephson, Tinkham}. By virtue of the oscillating current, the SJJ is generally discussed in terms of a rigid pendulum analogy. On the other hand, the BJJ dynamics resembles to that of a nonrigid pendulum with length-dependent angular momentum ($Z$)~\cite{hamiltonian}.

In the BJJ, the double-well trap is created by a laser field barrier that divides a single trapped condensate in two parts. Hence, the asymmetry of the wells ($\triangle E)$, as well as the barrier height (i.e. coupling) is controlled {\it externally} by the laser field. The coupling can further be modulated temporally by the laser beam.

To this end, we stress that the presence of $Z$-dependent nonlinear coupling in DWSP-JJ implies fundamentally different dynamical features from the BJJ and SJJ models, particularly in certain parameter regimes, as discussed in the next section.
\section{Results}
\label{sec:Results}
We start exploring the dynamical landscape of the DWSP-JJ model first by determining the stationary solutions at fixed points. Equation~(\ref{eq:DWSP_JJ_Z}) shows that the fixed points occur at ($\phi^*$=$2\pi n$) or ($\phi^*$=$(2n+1)\pi$) with the respective population imbalance values determined from
\begin{equation}
\ f(Z)|_{Z^*}= \Lambda Z^* + \triangle E+\frac{q(Z^*)Z^*}{\sqrt{1-(Z^*)^2}}\cos(\phi^*) = 0.
\label{eq:fz}
\end{equation}

We investigate the range $0.01 < \nu_p<0.35$, and $0.01< \eta<0.35$ including the resonant coupling ($\nu_p = 0.2, \eta = 0.2$)~\cite{proposal1}. For the scaled distance parameter $kd$ we use the range 3-12 and discuss the uncoupled limit ($kd >> 1$ qualitatively.\\

\subsection{Fixed Points}
\begin{figure}[h]
\begin{tabular}{ccc}
\epsfig{file=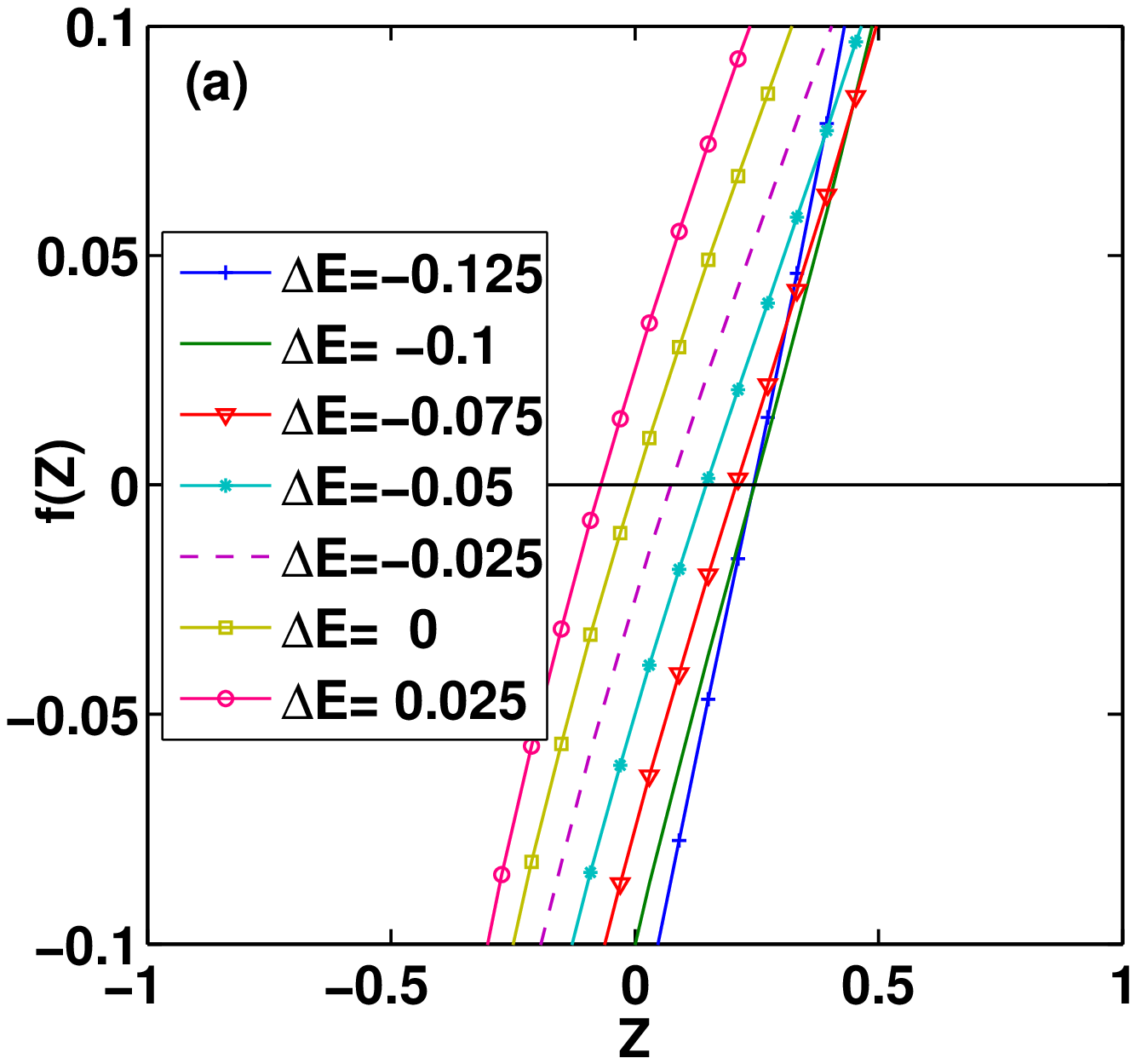,width=0.32\linewidth,clip=} &
\epsfig{file=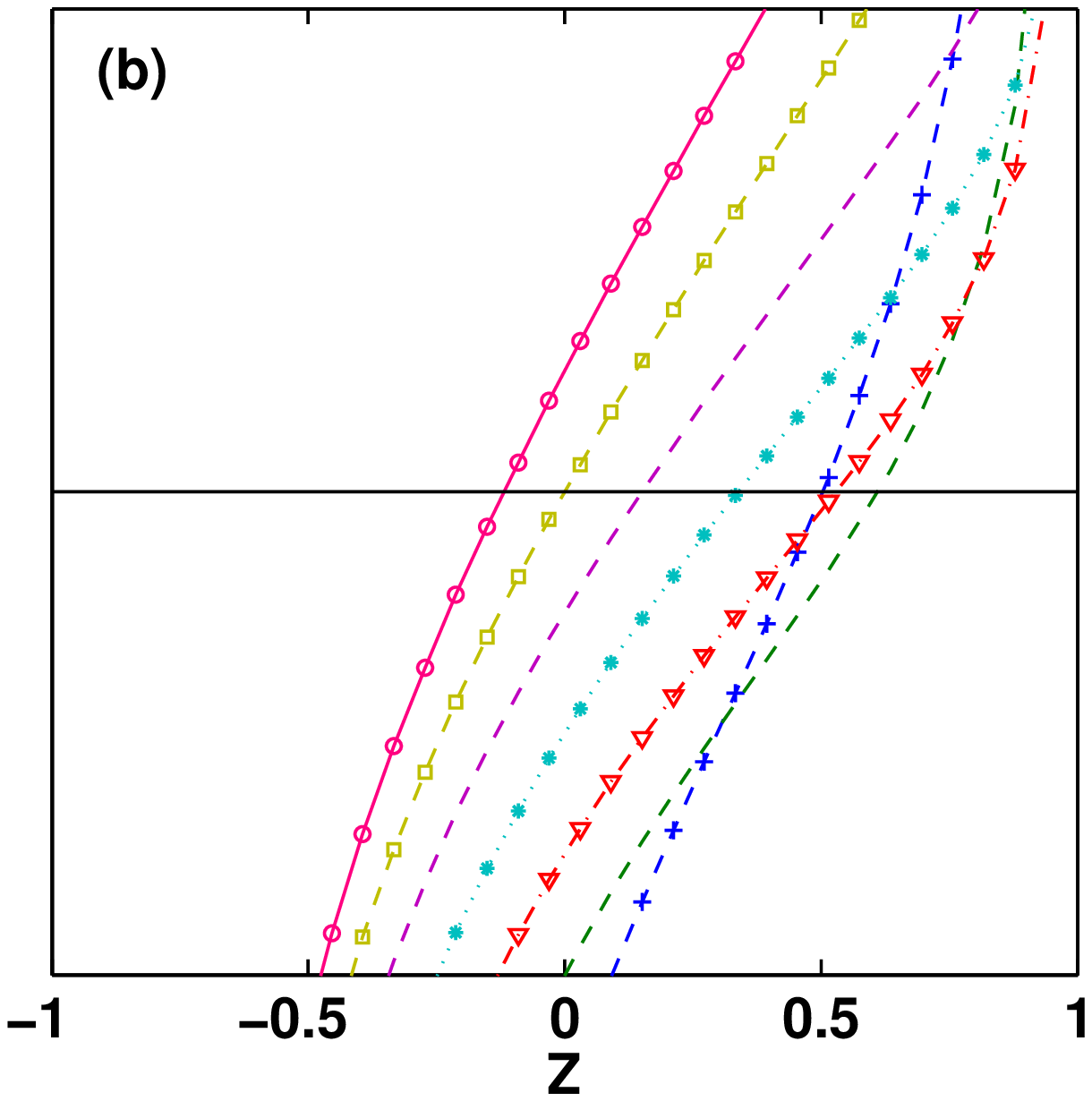,width=0.32\linewidth,clip=}
\epsfig{file=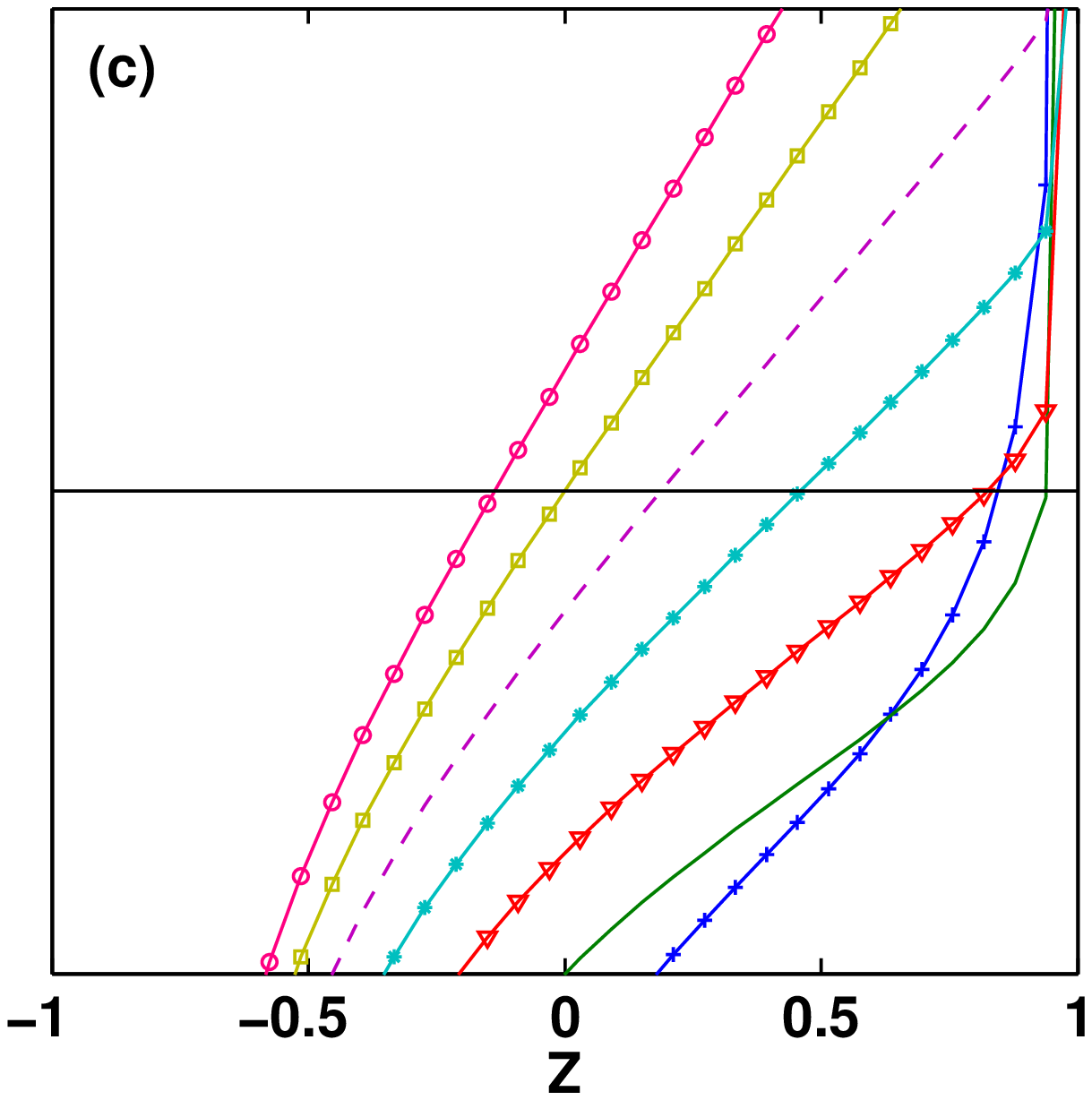,width=0.32\linewidth,clip=}
\end{tabular}
\caption{\scriptsize $f(Z)$ (Eq.~\ref{eq:fz}) for zero-phase modes plotted at $\nu_p = 0.15$ with various values of $\triangle E$ and (a) $kd = 3$ (b) $kd = 6$ and  (c) $kd = 12$ The critical points are the roots of $f(Z)=0$.}
\label{fig:1}
\end{figure}

The zero-phase modes describe the transfer of energy between the soliton and surface-plasmon states with zero time-average value of the phase. In figures~\ref{fig:1}(a-c) the f(Z) is plotted for different values of $\triangle E$ for $v_p = 0.15$ and $kd = 3, 6, 12$. For $\triangle E = 0$ the fixed point is at $Z = 0$, which can also be deduced by inspecting Eq.~(\ref{eq:fz}). This is analogous to the "symmetric double well" of the BJJ model.  For $\triangle E < (>) 0$, the fixed point occurs at nonzero fractional population imbalance $Z > (<) 0$.

\begin{figure}[h]
\begin{tabular}{ccc}
\epsfig{file=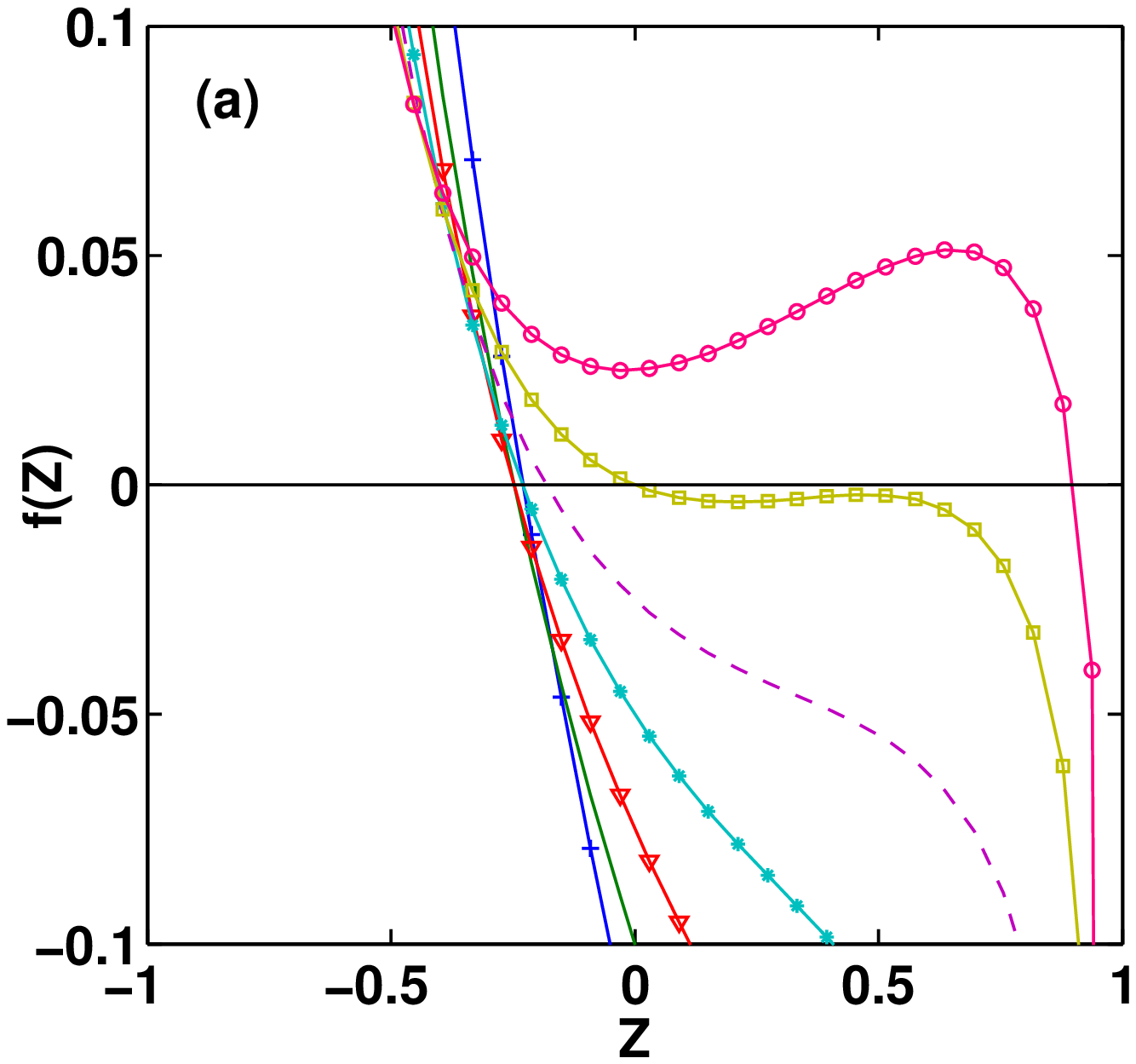,width=0.32\linewidth,clip=} &
\epsfig{file=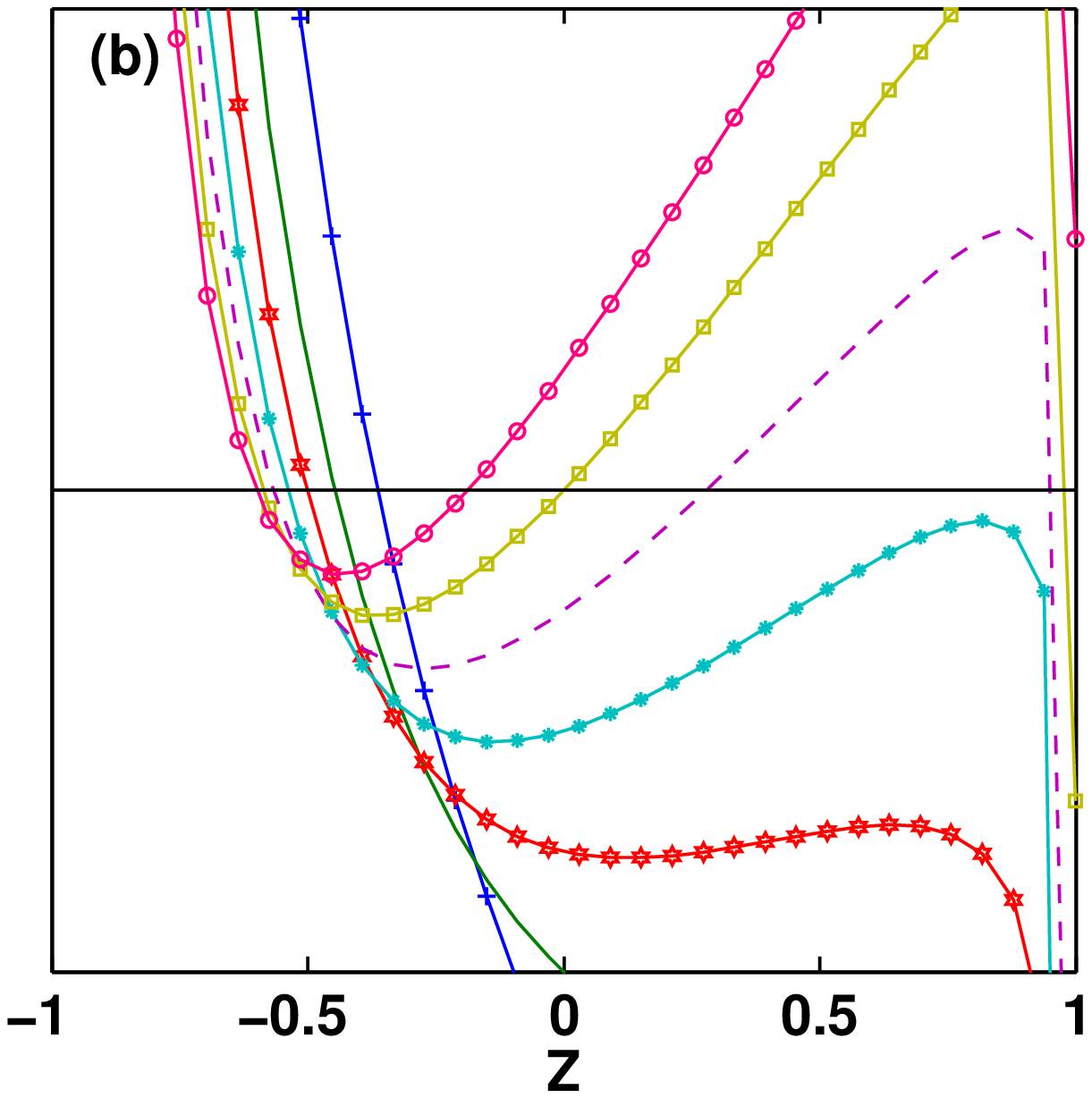,width=0.32\linewidth,clip=}
\epsfig{file=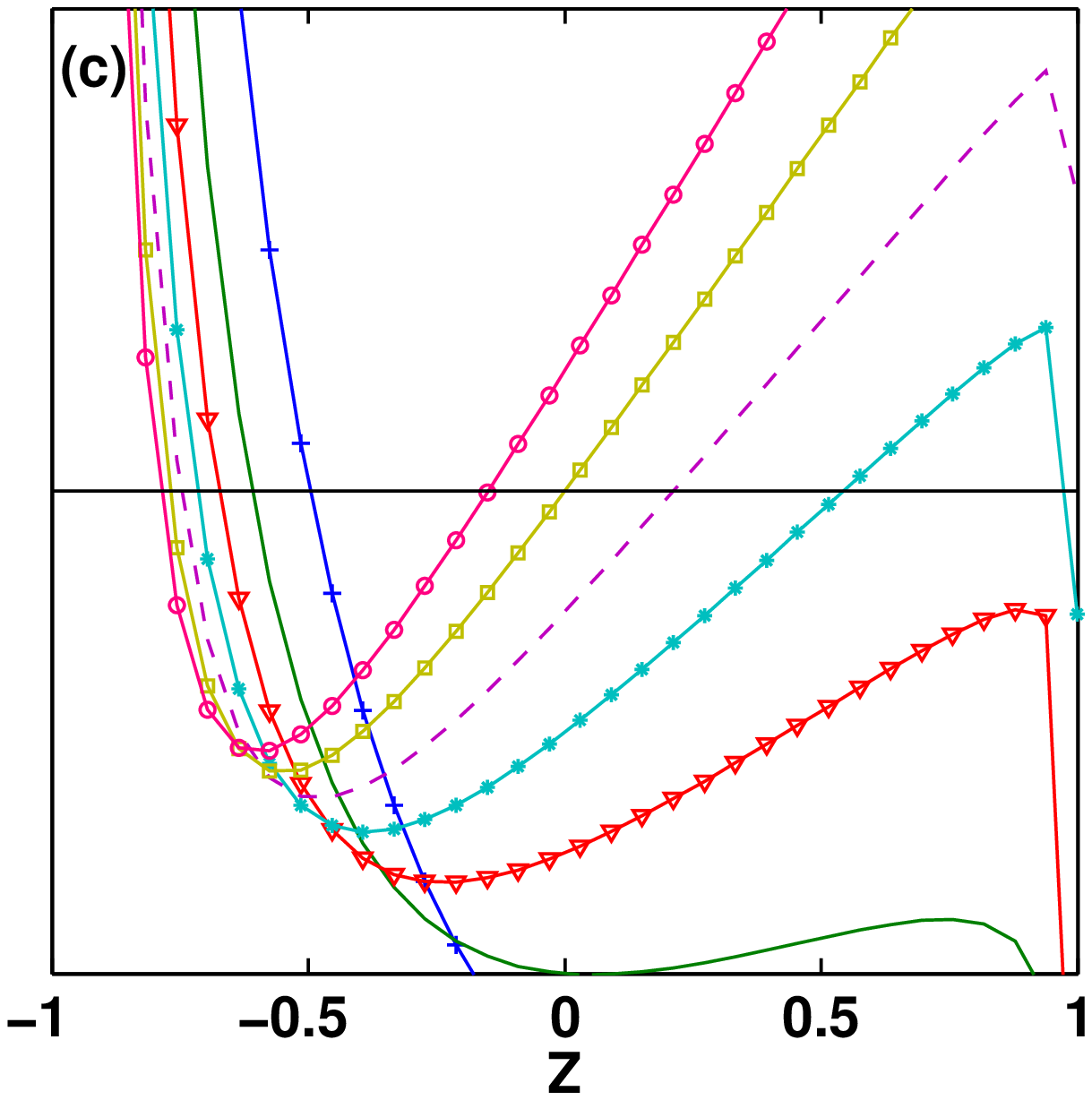,width=0.32\linewidth,clip=}
\end{tabular}
\caption{\scriptsize $f(Z)$ (Eq.~\ref{eq:fz}) for $\pi$-phase modes plotted at $\nu_p = 0.15$ with various values of $\triangle E$ and (a) $kd = 3$ (b) $kd = 6$ and  (c) $kd = 9$ The critical points are the root(s) of $f(Z)=0$.}
\label{fig:2}
\end{figure}

The other mode is the $\pi$-phase mode, where the time-average value of phase is $<\phi>=\pi$. Up to three fixed points emerge depending on the model parameters as shown in Fig.~(\ref{fig:2}). The location of the fixed points depend strongly on the scaled distance $kd$.\\


\subsection{Phase-space analysis}

\begin{figure}[h]
\epsfig{file=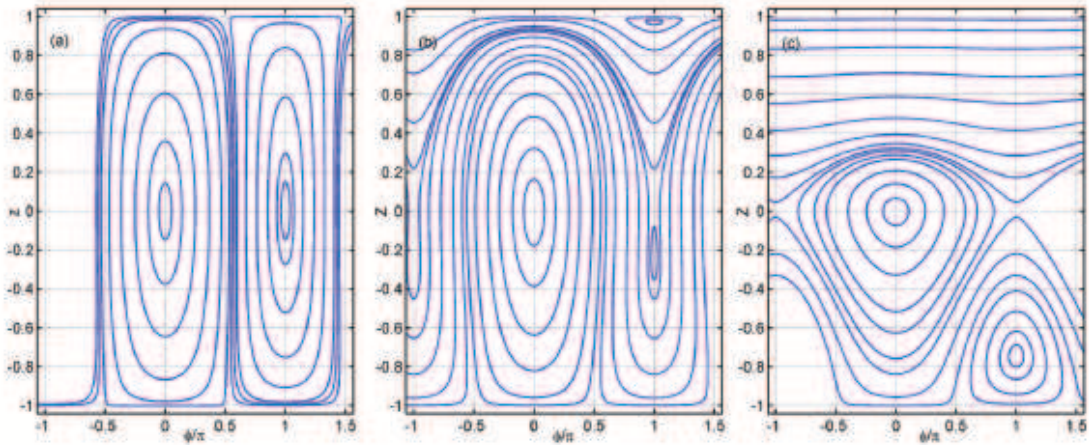,width=0.95\linewidth,clip=}
\caption{\scriptsize Several phase-space trajectories of DWSP-JJ system for $\triangle E = 0, \nu_p = 0.15$, (a)$kd = 3$, (b) $kd = 6$, (c) $kd = 12$.}
\label{fig:3}
\end{figure}

A concise description of the dynamical behavior can be given by the phase-space representation of
Eq.~\ref{eq:DWSP_JJ_Z}. In Figure~\ref{fig:3}(a-c), several phase space trajectories are given for the symmetric case with $\nu_p = 0.15$ and $kd$ = 3, 6, 12. For $kd = 3$ (Fig.~\ref{fig:3}(a)), the zero-phase and $\pi$-phase modes exhibit similar behavior, with rigid-pendulum type closed-orbit oscillations at small amplitudes and anharmonic closed-orbit oscillations at large amplitudes. The running phase trajectories consists of $|Z|\sim 1$ plateau connected by a population inversion occuring close to $\phi\sim(2n+1)\pi/2$ points. All these features are similar to the dynamical modes of the BJJ model discussed elsewhere~\cite{proposal10}.\\

When the spacing parameter is increased to $kd = 6$, we observe a drastic change in the phase-space portrait (Fig~\ref{fig:3}(b)). The anharmonicity of the large-amplitude zero-phase modes become prominent. For the $\pi$-phase, three fixed points appear, of which the $Z^*\sim 0.98$ and $Z^* \sim -0.25$ are enclosed by bounded trajectories, whereas the $Z^* = 0$ is an unstable fixed point. In Figure~\ref{fig:3}(c), we observe that increasing the $kd$ further ($kd = 12$) results in the decoupling of the system for soliton-dominant initial populations ($Z(0) > 0.5$). This is expected since the coupling parameter $q(Z)$ remains small. On the other hand, a surface-plasmon dominant initial population ($Z(0) < 0$) is still effectively "strong coupled" since the coupling parameter can be close to unity even though $kd$ is large (see Eq.~(\ref{eq:q})). The closed-orbits of the zero phase mode widens in phase towards the two unstable fixed points as opposed to being confined around $|\phi| < \pi/2$. The closed orbit $\pi$-phase modes occurs around $Z^* \sim -0.8$. In accordance with Fig.~\ref{fig:2}(c) another fixed point exists that occurs almost at $Z^* = 1$ and invisible in this scale.

\begin{figure}[h]
\epsfig{file=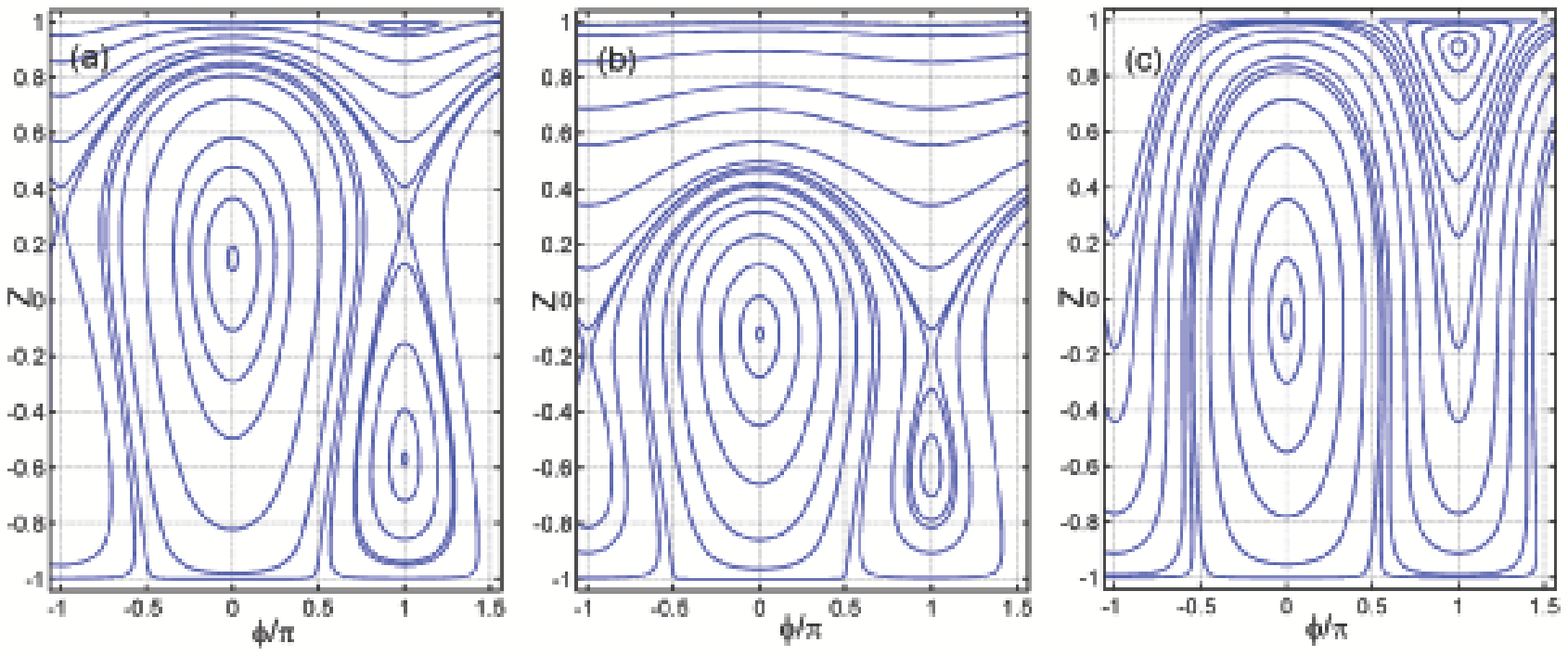,width=0.95\linewidth,clip=}
\caption{\scriptsize The phase-space trajectories of DWSP-JJ system for (a)$\triangle E = -0.025, \nu_p = 0.15, kd = 6$, (b) $\triangle E = 0.025, \nu_p = 0.15, kd = 6$, (c) $\triangle E = 0.025, \nu_p = 0.15, kd = 3$.}
\label{fig:4}
\end{figure}

The phase trajectories for ($\triangle E \neq 0$) and fixed $\nu_p = 0.15$ are given in Fig.~\ref{fig:4}. Figure~\ref{fig:4} (a) is plotted for $\triangle E = -0.025$, $kd = 6$. The zero-phase fixed point is located at $Z^* = 0.19$. The $\pi$-phase closed modes occur around $Z^* = -0.6$ and in a narrow region around $Z^* \sim 1$. For $\phi(0) = 0$ running phase orbits have large amplitude oscillations for $0.81 < Z < 0.85$ and small oscillations for $0.85 < Z$.
The phase diagram for $\triangle E = 0.025, kd = 6$ is shown in Fig~\ref{fig:4} (b). The zero phase mode is at ($Z^* = -0.2$). The $\pi$-phase mode has closed orbits only around ($Z^* = 0.95$) . When $kd$ is decreased ($kd = 3$) (Fig.\ref{fig:4}(c)) $\phi=\pi$ mode forms closed orbits around $Z^*=0.9$.

In the BJJ model, the parameter $\triangle E$ is a measure of the asymmetry between the two trapping states of atoms. Evidently, a nonzero $\triangle E$ induces a population imbalance between the two states. In the DWSP-JJ model, this asymmetry is indicated by the location of the $\phi^* = 0, Z^* \neq 0$ fixed points.

\begin{figure}[h]
\epsfig{file=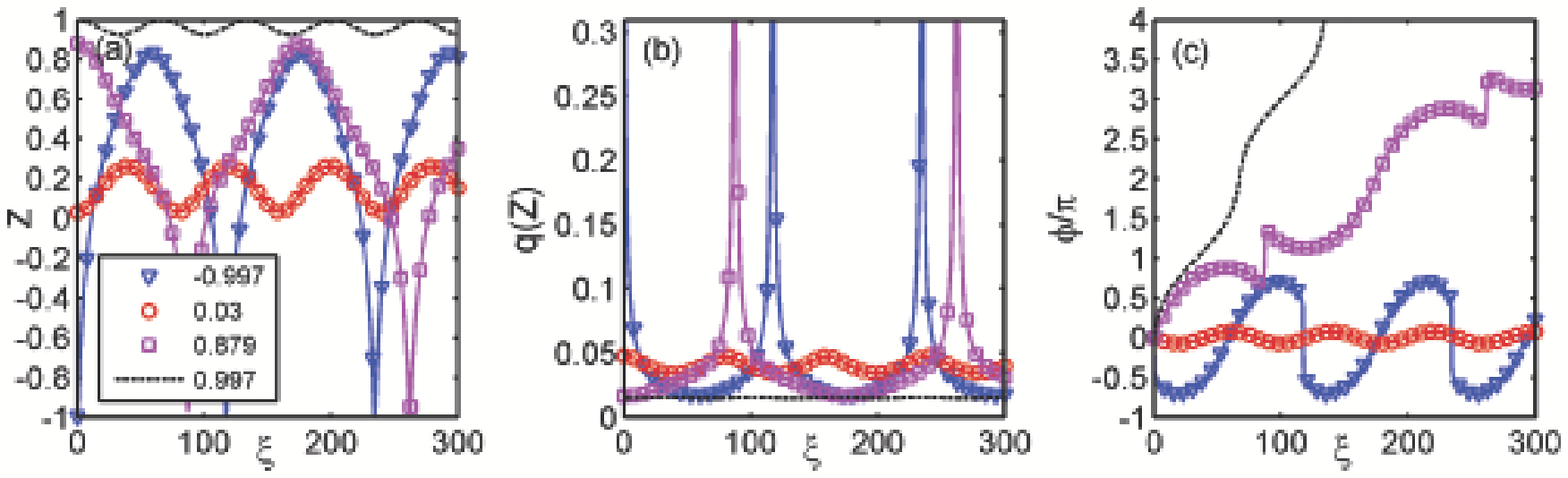,width=0.95\linewidth,clip=}
\centering
\caption{\scriptsize Propagation of (a) $Z$ (b) coupling parameter $q(Z)$, and (c) $\phi$ for $\phi(0)=0$ and for various values of $Z(0)$ indicated in the panel of (a). The model parameters are taken from Fig.~\ref{fig:4}(a)}
\label{fig:5}
\end{figure}

\begin{figure}[h]
\epsfig{file=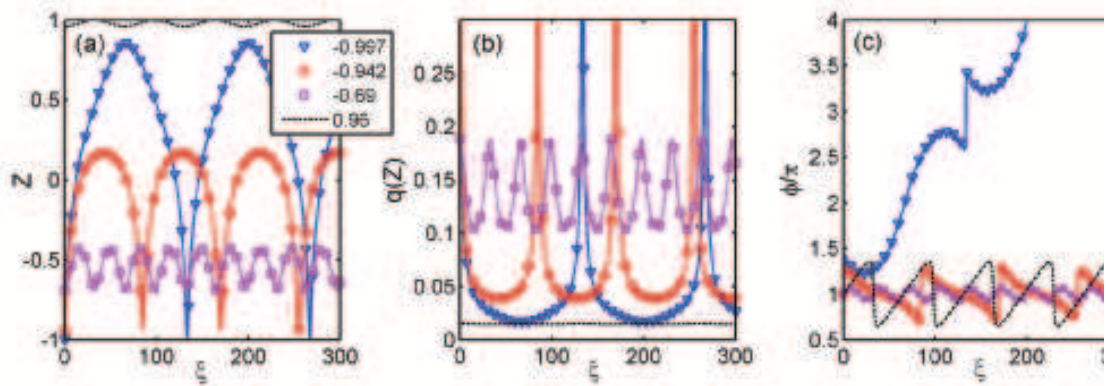,width=0.95\linewidth,clip=}
\caption{\scriptsize Propagation of (a) $Z$ (b) coupling parameter $q(Z)$, and (c) $\phi$ for $\phi(0)=\pi$ and for various values of $Z(0)$ indicated in the panel of (a). The model parameters are taken from Fig.~\ref{fig:4}(a)}
\label{fig:6}
\end{figure}

We next investigate the variation of $Z(\zeta)$, $q(Z)$, and $\phi(\zeta)$ for various trajectories from Fig~\ref{fig:4}(a). Figure~\ref{fig:5} show the propagation with initial relative phase $\phi(0) = 0$ for different initial values of $Z(0)$. As expected from the functional form in Eq.~(\ref{eq:q}), the coupling parameter shows variations commensurate with that of the population imbalance. Small amplitude oscillations (Fig.~\ref{fig:5}(a), circles and dotted line) have small, almost constant $q(Z)$ whereas large amplitude oscillations (triangle and square) exhibit impulsive behavior of the coupling. Similar observations are present for $\phi(0) = \pi$ in \ref{fig:6}.

\begin{figure}[h]
\epsfig{file=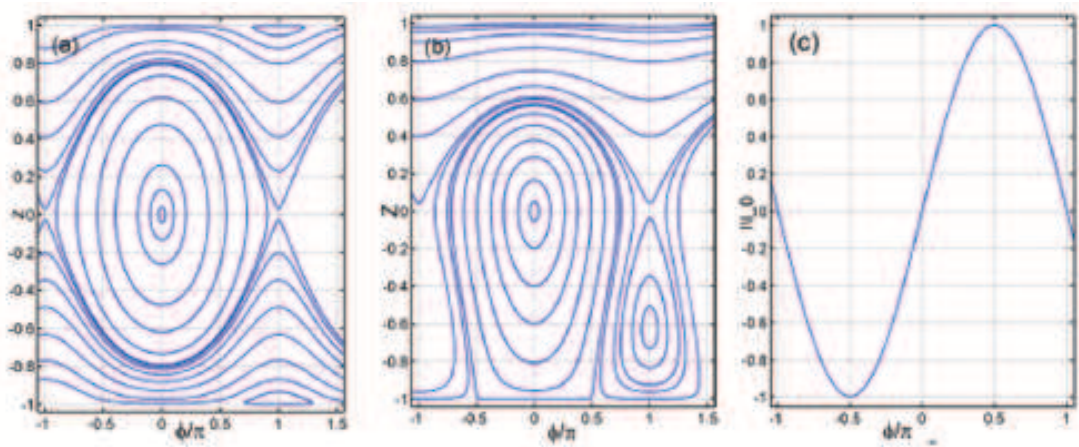,width=0.95\linewidth,clip=}
\caption{\scriptsize Comparison of the phase-space trajectories of (a) BJJ ($q\equiv 0.03, \triangle E = 0, \Lambda = 0.15$) and (b) DWSP-JJ ($kd = 6, \triangle E = 0, \Lambda = 0.15$) and (c) SJJ models}
\label{fig:7}
\end{figure}

We conclude our discussion by a qualitative comparison of the BJJ, DWSP-JJ and SJJ phase portraits in Fig.~\ref{fig:7} (a-c), respectively. The normalized SJJ AC current is plotted for a constant junction potential in Eq.~\ref{eq:SJJ}. A detailed comparison between SJJ and BJJ was reported in the literature~\cite{proposal10}. Here we rather focus on the BJJ and DWSP-JJ phase space features.

The BJJ model is depicted with an effective coupling constant of $q\sim0.03$ which is within the dynamic range of the $q(Z)$ of DWSP-JJ. The phase space of BJJ is decorated with open- and bounded-phase trajectories that are symmetric with respect to the $Z$ axis. The $\phi(0) = 0$ trajectories are bounded for $|Z(0)| < 0.8$. The bounded orbits are harmonic for $|Z(0)| < 0.5$ and become anharmonic for $0.5 < Z(0) < 0.8$. The $\phi(0) = \pi$ trajectories are open for $|Z(0)| < 0.9$ and become bounded anharmonic orbits for $0.9 < |Z(0)| < 1$.

In the DWSP-JJ phase portrait plotted with a comparable parameter set (Fig.~\ref{fig:7}(b)), the asymmetry in $Z$ induced by the dynamical coupling parameter is prominent. Qualitatively speaking, the DWSP-JJ acts like a double-well trap BJJ system, where the barrier height (i.e. the coupling) between the two wells depends on the population of one well (i.e. the soliton amplitude).  The bound trajectories are strongly anharmonic except for $|Z(0) < 0.2|$. For $\phi(0) = 0$, the orbits are closed for all negative $Z(0)$ values and for positive $Z(0) < 0.55$. The $\phi(0) = \pi$ states are bounded for $-0.9 < Z(0) < 0$.

Finally, we would like to point out the dissipative effects that are excluded from the present dynamical model. Naturally, both population- and phase-dependent dissipation mechanisms are applicable to this model. The effects of phase dissipation are discussed in part in~\cite{proposal1}. We plan to discuss a dissipative-DWSP-JJ model in a subsequent work. In particular, the absorption loss of the surface-plasmon will be relevant and may alter the phase portrait substantially for strong loss. Nevertheless, the qualitative features of the present analysis are still providing insight and reveal intriguing properties that can be observable under weak dissipative conditions.

\section{Conclusion}
\label{sec:Conclusion}
In this work, we proposed a weakly coupled optical-soliton and metal surface-plasmon system as a novel type of Josephson Junction and investigated its dynamical properties. We have found that the coupling parameter that depends on the soliton amplitude allows rich dynamical features, different than that observed in bosonic Josephson Junctions. The DWSP-JJ is potentially a convenient system to investigate collective dynamics of photons and surface plasmons, that may reveal novel classical and quantum plasmonic phenomena.

\ack
K. G\"{u}ven acknowledges the support by the Turkish Academy of Sciences.

\end{document}